\documentclass[12pt]{article}

\usepackage[latin1]{inputenc}
\usepackage{amsmath}
\usepackage{amsfonts}
\usepackage{amssymb}
\usepackage[dvips]{graphicx}
\usepackage{sectsty}
\usepackage{a4wide}

\usepackage{color}

\definecolor{blue}{rgb}{0,0,1}

\definecolor{green}{rgb}{0,1,0}

\definecolor{red}{rgb}{1,0,0}

\allsectionsfont{\fontseries{m}\sffamily}

\begin{document}
\author{Holger Gies and Klaus Klingm\"uller
\\
\\
{\small Institute for Theoretical Physics, Heidelberg University}\\
{\small Philosophenweg 16, D-69120 Heidelberg, Germany}\\
{\small h.gies@thphys.uni-heidelberg.de,
  k.klingmueller@thphys.uni-heidelberg.de} }
\title{Worldline algorithms for Casimir configurations}
\begin{sffamily}
\maketitle
\begin{abstract}
  \noindent
  \normalfont We present improved worldline numerical algorithms for
  high-pre\-cision calculations of Casimir interaction energies
  induced by scalar-field fluctuations with Dirichlet boundary
  conditions for various Casimir geometries. Significant reduction of
  numerical cost is gained by exploiting the symmetries of the
  worldline ensemble in combination with those of the configurations.
  This facilitates high-precision calculations on standard PCs or
  small clusters. We illustrate our strategies using the
  experimentally most relevant sphere-plate and cylinder-plate
  configuration. We compute Casimir curvature effects for a wide
  parameter range, revealing the tight validity bounds of the commonly
  used proximity force approximation (PFA). We conclude that data
  analysis of future experiments aiming at a precision of 0.1\% must
  no longer be based on the PFA. Revisiting the parallel-plate
  configuration, we find a mapping between the $D$-dimensional Casimir
  energy and properties of a random-chain polymer ensemble.
\end{abstract}
\end{sffamily}

\section{Introduction}

Recent years have witnessed remarkable qualitative and quantitative
progress in the understanding of the Casimir effect \cite{Casimir:dh}
both experimentally and theoretically.  Measurements of the Casimir
force have reached a precision level of 1\%
\cite{Lamoreaux:1996wh,Mohideen:1998iz,Roy:1999dx,Ederth:2000,%
Chan:2001,Chen:2002}.  Further improvements are currently aimed at
with intense efforts, owing to the increasing relevance of these
quantum forces for nano- and micro-scale mechanical systems. Also from
the perspective of fundamental physics, Casimir precision measurements
play a major role in the search for new sub-millimeter forces,
resulting in important constraints for new physics
\cite{Bordag:1998nv,Long:1998dk,%
Mostepanenko:2001fx,Milton:2001np,Decca:2003td}.

On this level of precision, corrections owing to surface roughness,
finite conductivity, thermal fluctuations and geometry dependencies
have to be accounted for
\cite{Klimtchiskaya:1999,Lambrecht:1999vd,Bezerra:2000,%
  Bordag:2001qi,Lambrecht:2005}.  These corrections may be classified
in terms of material corrections on the one hand; they are induced,
for instance, by surface roughness and finite conductivity which may
be viewed as a deviation from the ideal Casimir configuration. On the
other hand, corrections due to geometry dependence are of direct
quantum origin and thus universal, i.e., independent of the
microscopic details of the interactions between the fluctuating field
and the constituents of the surfaces.  Since material corrections are
difficult to control with high precision, force measurements at larger
surface separations up to the micron range are
intended.\footnote{Measurements at larger surface separations are also
  aimed at in order to resolve a recent controversy about thermal
  corrections, see \cite{Mostepanenko:2005qh,Brevik:2006jw} and
  references therein.  Even though thermal contributions are also
  universal in the {\em ideal} Casimir limit, they can mix
  nontrivially with material corrections in a way that may affect any
  {\em real} experiment \cite{Sernelius}.} Though this implies
stronger geometry dependence, this universal effect is, in principle,
under clean theoretical control, since it follows directly from
quantum field theory \cite{Graham:2002xq}.

Straightforward computations of geometry dependencies have long been
conceptually complicated, since the relevant information is subtly
encoded in the fluctuation spectrum.  Generically, analytic solutions
are restricted to highly symmetric geometries. This problem is
particularly prominent, since current and future precision
measurements predominantly rely on configurations involving curved
surfaces, such as a sphere above a plate. Curved configurations help
to circumvent the difficulty of maintaining parallelism as it occurs
in the parallel-plate configuration; the latter has been mastered so
far only in one experiment \cite{Bressi:2002fr} with a precision level
of $\sim 15\%$. As a general recipe for curved configurations, the
proximity force approximation (PFA) \cite{pft1} has been the standard,
though uncontrolled, tool for estimating curvature effects for
non-planar geometries in all experiments so far.

In recent years, various new techniques have been developed for
computing Casimir effects in more involved geometries
\cite{semicl,Golestanian:1998bx,Graham:2002xq,Gies:2003cv,%
  Scardicchio:2004fy,Bulgac:2005ku,Emig:2006uh,Bordag:2006vc}, each with 
its own merits and limitations.  This includes improved approximation
techniques which can deal with curved geometries more reliably, such
as the semiclassical approximation \cite{semicl} and the optical
approximation \cite{Scardicchio:2004fy}, as well as exact field
theoretic methods based on functional-integral techniques
\cite{Golestanian:1998bx,Emig:2006uh,Bordag:2006vc} or scattering theory
\cite{Graham:2002xq,Bulgac:2005ku}.

In this work, we use and further develop {\em worldline numerics}
\cite{Gies:2003cv,Gies:2001zp}, which facilitates Casimir computations
from field-theoretic first principles. Worldline numerics builds on a
combination of the string-inspired approach to quantum field theory
\cite{Schubert:2001he} with Monte Carlo methods. As a main advantage,
the worldline algorithm can be formulated for arbitrary geometries,
resulting in a numerical estimate of the exact answer
\cite{Gies:2003cv}. The inherent use of Feynman path-integral
techniques circumvents the problem of determining the Casimir
fluctuation spectrum \cite{Gies:2005ym}, which is often encountered in
other approaches.  The resulting algorithms are trivially scalable and
computational efforts increases only linearly with the parameters of
the numerics.

Here, we present worldline algorithms to examine the Casimir effect in
a sphere-plate and cylinder-plate geometry for a fluctuating scalar
field, obeying Dirichlet boundary conditions (``Dirichlet scalar'').
We compute the Casimir interaction energies that give rise to forces
between the rigid surfaces.  This allows for a quantitative
determination of validity bounds of approximation methods such as the
PFA, some results of which have already been presented in a recent
Letter \cite{Gies:2006bt}. We detail significant improvements of the
numerical algorithms which facilitate high-precision calculations.
Apart from numerical discretization errors which are kept at or below
the 0.1\% level, no quantum-field-theoretic approximation is needed.
Our results further strengthen the agreement with recently obtained
analytic solutions for medium or larger curvature
\cite{Bulgac:2005ku,Emig:2006uh} and for small curvature
\cite{Bordag:2006vc}.

We emphasize that the Casimir energies for the Dirichlet scalar should
generally not be taken as an estimate for those for the
electromagnetic (EM) field, leaving especially the experimentally most
used sphere-plate case as a pressing open problem. Nevertheless, a
comparison with other techniques can meaningfully be performed, and
the validity constraints that we derive, e.g., for the PFA hold
independently of the type of boundary condition, since the PFA
approach makes no reference to the nature of the fluctuating field.
If an experiment is performed outside the PFA validity ranges
determined below, any comparison of the data with theory using the PFA
has no firm basis. 

We also revisit Casimir's classic parallel-plate configuration, first
because further algorithmic strategies can easily be illustrated here;
and second, we thereby obtain a mapping between the $D$ dimensional
Casimir effect and characteristic properties of a random-chain polymer
ensemble (say, in 3 space dimensions), due to the use of path
integrals in the worldline method. 

Our work is organized as follows: in Sect.~\ref{sec:CEW}, we briefly
review elements of the worldline formulation for Casimir
configurations as well as the basic ideas of worldline numerics.  In
Section \ref{sec:WACC}, the construction of our new worldline
algorithms is detailed for various Casimir configurations. We present
our conclusions in Sect. \ref{sec:C}. We close this introduction with
a brief review of Casimir curvature effects and the PFA; the latter is
not only a simple (though potentially misleading) approximation, but
also provides for a useful normalization for our numerical results.

\subsection{Casimir curvature effects and proximity force
  approximation (PFA)} 

An intriguing property of the Casimir effect has always been its
geometry dependence. As long as the typical curvature radii $R_i$ of
the surfaces are large compared to the surface separation $a$, the PFA
is assumed to provide for a good approximation. In this approach, the
curved surfaces are viewed as a superposition of infinitesimal
parallel plates \cite{pft1,Bordag:2001qi}. The Casimir interaction
energy is obtained by an integration of the parallel-plate energy
applied to the infinitesimal elements. Part of the curvature effect is
introduced by the choice of a suitable integration measure which is
generally ambiguous, as discussed, e.g., in \cite{Scardicchio:2004fy}.
For the case of a sphere with radius $R$ at a (minimal) distance $a$
from a plate, the PFA result at next-to-leading order reads
\begin{eqnarray}
\text{}\!\!\!E_{\text{PFA}}(a,R)\!
&=&\! E_{\text{PFA}}^{(0)}(a,R)\left(\! 1-
\genfrac{\{}{\}}{0pt}{}{1}{3} \frac{a}{R}  
+ \mathcal O ((\tfrac{a}{R})^2)\!\! \right)\!,
\label{eq:Enorm} \\
&&E_{\text{PFA}}^{(0)}(a,R)=-c_{\text{PP}}\frac{\pi^3}{1440}
\frac{R}{a^2}, \label{eq:Ezero}
\end{eqnarray}
where the upper (lower) coefficient in braces holds for the so-called
plate-based (sphere-based) PFA. They represent two limiting cases of
the PFA and have often been assumed to span the error bars for the
true result. Furthermore, $c_{\text{PP}}=2$ for an EM field or a
complex scalar, and $c_{\text{PP}}=1$ for a real scalar field.

The first field-theoretic confirmation of the zeroth-order result
$E_{\text{PFA}}^{(0)}(a,R)$ has been obtained within the
semi-classical approximation in \cite{semicl}. We will use this
zeroth-order interaction energy (and its analogue for the
cylinder-plate configuration) as a normalizer for our numerical
estimates. As an advantage, any deviation from this result can be
interpreted as a true quantum-induced Casimir curvature effect. Future
experiments are indeed expected to become sensitive to the first-order
curvature correction, which therefore is of particular interest to us.

Conceptually, the PFA is in contradiction with Heisenberg's
uncertainty principle, since the quantum fluctuations are assumed to
probe the surfaces only locally at each infinitesimal element.
However, fluctuations are not localizable, but at least probe the
surface in a whole neighborhood. In this manner, the curvature
information enters the fluctuation spectrum.  This quantum mechanism
is immediately visible in the worldline formulation of the Casimir
problem. Therein, the sum over fluctuations is mapped onto a Feynman
path integral, see below. Each path (worldline) can be viewed as a
random spacetime trajectory of a quantum fluctuation. Owing to a
generic spatial extent of the worldlines, the path integral directly
samples the curvature properties of the surfaces \cite{Gies:2003cv}.

\section{Casimir effect on the worldline}
\label{sec:CEW}

\subsection{Worldline formulation for a Dirichlet scalar}

Let us briefly recall from \cite{Gies:2003cv} how the Casimir effect
for a real scalar field $\phi$ satisfying Dirichlet boundary
conditions can be described in the worldline formalism.  For this, the
field is coupled to a background potential $V(x)\geq0$ which models
the Casimir configuration: qualitatively, the amplitude of the $\phi$
field fluctuations are suppressed at those regions of spacetime where
the potential $V(x)$ is large. The field-theoretic Euclidean Lagrangian is
\begin{equation}
\mathcal L=\frac{1}{2}\partial_\mu\phi\partial_\mu\phi
+\frac{1}{2}m^2\phi^2+\frac{1}{2}V(x)\phi^2.
\end{equation}
Here, we have included a mass term for the scalar; the massless limit,
which is more analogous to the photon field, can always safely be
taken.  From the standard generating functional for the quantum
correlation functions of $V(x)$, we obtain the (unrenormalized)
effective action, which reads in worldline representation in $D=d+1$
dimensional Euclidean spacetime:
\begin{equation}
\Gamma[V]=-\frac{1}{2}\frac{1}{(4\pi)^{D/2}}
\int_0^\infty\frac{dT}{T^{1+D/2}}e^{-m^2T}
\int d^Dx_\mathrm{CM}
\bigl\langle e^{-\int_0^Td\tau V(x(\tau))}-1\bigr\rangle_x.
\label{eq:EA}
\end{equation}
The expectation value $\langle \dots \rangle_x$ has to be taken with
respect to an ensemble of closed worldlines with Gau\ss ian velocity
distribution,
\begin{equation}
\langle \dots \rangle_x
:=\int_{x(T)=x(0),\atop\mathrm{CM}} \mathcal D x \, \dots
e^{-\frac{1}{4} \int_0^T d \tau {\dot x}^2}
\label{eq:VEV2}
\end{equation}
with implicit normalization $\langle 1 \rangle_x=1$. For convenience,
the common center of mass $x_\mathrm{CM}$ of the worldlines
has been separated off.

In this work, we concentrate on Casimir forces between disconnected
rigid bodies which we represent by a time-independent potential
$V(\mathbf x)=V_1(\mathbf x)+V_2(\mathbf x)+\cdots$; the potentials
$V_i(\mathbf x)$ for the single bodies have pairwise disjoint
supports, i.e., $V_i(\mathbf x)V_j(\mathbf x)=0$ for all $\mathbf x$
and $i\neq j$.  From the effective action, we obtain the Casimir
energy by scaling out the trivial time integration,
\begin{equation}
\mathcal E=\frac{\Gamma}{\int dx_{0,\text{CM}}}.
\label{eq:CE}
\end{equation}
For the Casimir force, only the portion of the Casimir energy which
depends on the relative position of the objects is relevant. This
portion can conveniently be extracted from the total Casimir energy by
subtracting the (self-)energies of the single objects. This leads us
to the \emph{Casimir interaction energy},
\begin{equation}
E_{\text{Casimir}}:=\mathcal E_{V_1+V_2+\cdots}
-\mathcal E_{V_1}-\mathcal E_{V_2}
-\cdots,
\label{eq:CIE}
\end{equation}
which serves as the potential energy for the Casimir force; i.e.,
Casimir forces (or torques, etc.) are obtained by the (negative)
derivative of $E_{\text{Casimir}}$ with respect to a distance (or
angle) parameter.  By this procedure, also any UV divergencies of Eq.
\eqref{eq:EA} are automatically removed. Moreover, the interaction
energy can thus be well defined even if the Casimir (self-)energy of a
single surface is ill-defined in the ideal boundary-condition limit
(``perfect conductivity'', infinitely thin surfaces, etc.)
\cite{Deutsch:1978sc,Barton:2001wd,Graham:2003ib}.

For the ideal limit of infinitely thin surfaces, the potential
$V(\mathbf x)$ becomes a $\delta$ function in space,
\begin{equation}
V(\mathbf x)=\lambda\int_\Sigma d\sigma\ \delta^{(d)}(\mathbf x-\mathbf
x_\sigma). 
\label{eq:pot}
\end{equation}
The geometry of the Casimir configuration is defined by $\Sigma$,
denoting a $d-1$ dimensional surface. The surface measure $d\sigma$ is
assumed to be re-parameterization invariant, and $\mathbf x_\sigma$
denotes a vector pointing onto the surface. For a typical
configuration, $\Sigma$ consists of two disconnected objects (e.g.,
two disconnected plates), $\Sigma=\Sigma_1\cup\Sigma_2$, with
$\Sigma_1\cap\Sigma_2=\emptyset$. The positive coupling $\lambda$ has mass
dimension 1. In the ideal limit $\lambda\to\infty$, the potential
imposes Dirichlet boundary conditions on the quantum field.

For the potential Eq. \eqref{eq:pot}, the $\tau$ integral in the
expectation value in Eq. \eqref{eq:EA} reads
\begin{equation}
I_{\Sigma}[\mathbf x(\tau)]
:=\int_0^Td\tau\ V(\mathbf x(\tau))
=\lambda\sum_{\{\tau_\sigma:\mathbf x(\tau_\sigma)\in\Sigma\}}
\frac{1}{|\dot {\mathbf x}_\perp(\tau_\sigma)|}\ ,
\end{equation}
where the sum goes over all intersection points of the worldline
$x(\tau_\sigma)$ and the surface $\Sigma$. In the denominator, $\dot
{\mathbf x}_\perp(\tau_\sigma)$ denotes the component of the $\tau$
derivative perpendicular to the surface.

Computing the Casimir interaction energy Eq. \eqref{eq:CIE} for two
surfaces $\Sigma_1$ and $\Sigma_2$, the argument of the expectation
value in \eqref{eq:EA} becomes
\begin{equation}
\left(e^{-I_{\Sigma_1\cup\Sigma_2}[\mathbf x(\tau)]}-1\right)
-\left(e^{-I_{\Sigma_1}[\mathbf x(\tau)]}-1\right)
-\left(e^{-I_{\Sigma_2}[\mathbf x(\tau)]}-1\right)
\in[0,1].
\label{eq:Iint}
\end{equation}
Most importantly, Eq.~\eqref{eq:Iint} is nonzero only if the loop
$\mathbf x(\tau)$ intersects both surfaces.  In the Dirichlet limit
$\lambda\to\infty$, this expression then equals one. Thus, for a
massless scalar field with Dirichlet boundaries in $D=3+1$, the
worldline representation of the Casimir interaction energy boils down
to \cite{Gies:2003cv,Gies:2005ym}
\begin{equation}
E_{\text{Casimir}}=-\frac{1}{2} \frac{1}{(4\pi)^2} \int_{0}^\infty
\frac{d T}{T^3}\int d^3x_\mathrm{CM} 
\, \left\langle\Theta_\Sigma[\mathbf x(\tau)]
\right\rangle_{\mathbf x} . \label{eq:ECasW} 
\end{equation}
Here, the worldline functional $\Theta_\Sigma[\mathbf x(\tau)]=1$ if
the path $\mathbf x(\tau)$ intersects the surface
$\Sigma=\Sigma_1\cup\Sigma_2$ in both parts $\Sigma_1$ and $\Sigma_2$,
and $\Theta_\Sigma[\mathbf x(\tau)]=0$ otherwise, analogous to the
standard step function.

This compact formula has an intuitive interpretation: the worldlines
can be viewed as the spacetime trajectories of the quantum
fluctuations of the $\phi$ field. Any worldline that intersects the
surfaces does not satisfy Dirichlet boundary conditions. All
worldlines that intersect both surfaces thus should be removed from
the ensemble of allowed fluctuations, thereby contributing to the
negative Casimir interaction energy. The auxiliary integration
parameter $T$, the so-called propertime, effectively governs the
extent of a worldline in spacetime. Large $T$ correspond to IR
fluctuations with large worldlines, small $T$ to UV fluctuations.
Those $T$ values for which the spatial extent of the worldlines is
just big enough to intersect with both surfaces generically dominate
the Casimir interaction energy. Within the worldline picture, it is
already intuitively clear that for generic surfaces at a (suitably
defined\footnote{A useful definition may be given by the following
  construction: let $a>0$ be the maximally possible distance between two
  auxiliary parallel plates that can be placed in between the surfaces
  constituting the Casimir configuration without mutual intersection.
  This excludes pathological cases such as surfaces which are folded
  into each other. In this construction, it should also be understood
  that a change of $a$ should not be accompanied by a rotation of one
  of the surfaces.})  distance $a$ the Casimir interaction energy for
a Dirichlet scalar is negative and a monotonously increasing function
of $a$; therefore, the resulting force is always attractive in
agreement with a recent theorem \cite{Kenneth:2006vr}.

\subsection{Worldline numerics}

For the numerical evaluation of the expectation value
Eq.~\eqref{eq:VEV2}, two discretizations are required: first, the path
integral is approximated by a finite sum over an ensemble of
$n_\mathrm L$ random paths $\mathbf x_\ell(\tau),\
\ell=1,\dots,n_\mathrm L$, each of them forming a closed loop in
space(-time). Second, the propertime which parameterizes each path is
discretized:
\begin{equation}
\mathbf x_\ell(\tau),\ \tau\in[0,T]\,\,\longrightarrow \,\,
\mathbf x_{\ell k}
:=\mathbf x_\ell(k\cdot T/N),\ k=1,\dots,N; 
\end{equation}
i.e., the paths themselves are represented by $N$ points per loop
(ppl). Thus, the ensemble is described by a two dimensional array of
space vectors $(\mathbf x_{\ell k})$, with the indices $\ell$ and $k$
specifying the loop and the point on the loop, respectively.

We generate the random paths using the \emph{v-loop} algorithm
\cite{Gies:2003cv}. This algorithm incorporates the Gaussian term
$e^{-\frac{1}{4} \int_0^T d \tau {\dot {\mathbf x}}^2}$ as probability
distribution, so that the path integral in Eq. (\ref{eq:VEV2})
becomes an arithmetic mean:
\begin{equation}
\int_{\mathbf x(T)=\mathbf x(0),\atop\mathrm{CM}}
\mathcal D \mathbf x \, (\dots)
e^{-\frac{1}{4} \int_0^T d \tau {\dot {\mathbf x}}^2}
\,\,\longrightarrow\,\,
{\frac{1}{ n_\mathrm{L}}}\sum_{\ell=1}^{n_\mathrm{L}}(\dots)\label{eq:mean}.
\end{equation}
It is sufficient to generate only one so called \emph{unit-loop}
ensemble $(\mathbf y_{\ell k})$, i.e., an ensemble of loops with center of mass
$\mathbf x_\mathrm{CM}=0$ and $T=1$. An ensemble with other values for
$\mathbf x_\mathrm{CM}$ and $T$ is then simply obtained by computing
\begin{equation}
\mathbf x_{\ell k}=\mathbf x_\mathrm{CM}+\sqrt T\mathbf y_{\ell k}\label{eq:lt}
\end{equation}
for all $\ell$ and $k$.
At the same time, this technique provides for an analytic knowledge
of the integrand's $T$ dependence, which can be utilized for the $T$
integration.

With this discretization, the Casimir interaction energy Eq.
(\ref{eq:ECasW}) reads
\begin{equation}
E_{\text{Casimir}}=-\frac{1}{2}\frac{1}{(4\pi)^2}\int_0^\infty\frac{dT}{ T^3}
\int d^3x_\mathrm{CM}\frac{1}{n_\mathrm L}\sum_{\ell=1}^{n_\mathrm{L}}
\Theta_\Sigma[\mathbf x_\mathrm{CM}+\sqrt T\mathbf y_\ell]
\label{eq:ECasWd}.
\end{equation}
The discretization error is controlled by the two parameters
$n_\mathrm{L}$ and $N$. The number of loops per ensemble
$n_\mathrm{L}$ is related to a statistical error of the arithmetic
mean in Eq. (\ref{eq:mean}), which can be determined by jack-knife
analysis. The number of points per loop $N$ is chosen sufficiently
large to achieve the desired precision by studying the convergence of
the result. In this work, we have used ensembles with up to
$n_{\text{L}}=2.5\cdot10^5$ and $N=4\cdot10^6$.

The major advantages of worldline numerics are its scalability and its
independence of the background (geometry). The computational effort
scales only linearly with the parameters $n_{\text{L}}$, $N$, $D$,
etc, and the algorithm can be formulated for any given background
geometry. A disadvantage is that the statistical error decreases only
with $1/\sqrt{n_{\text{L}}}$, as for any Monte-Carlo method. This
implies that high-precision computations may require high statistics,
in contrast to estimates with, say, a few-percent error which require
very little computational effort. 

For high-precision Casimir applications with an intended error of
$\lesssim 0.1$\%, the CPU time needed for the evaluation of Eq.
(\ref{eq:ECasWd}) can be reduced substantially by specializing the
algorithm to the given Casimir geometry. Although this corresponds to
a loss of generality, we believe that the strategies which we describe
in the following, e.g., for the sphere-plate configuration, are
examples for a general set of algorithmic tools which will be useful
also for other Casimir configurations.

\section{Worldline algorithms for Casimir configurations}
\label{sec:WACC}

The general structure of a worldline algorithm for computing Casimir
interaction energies is summarized by Eq.~\eqref{eq:ECasWd}. The only
part of the algorithm that depends on the geometry consists of a
diagnostic routine which checks whether a given loop (for given
$\mathbf x_{\text{CM}}$ and $T$ intersects with (more than one of) the
surfaces $\Sigma_1,\Sigma_2,\dots$. The result of this diagnostic
routine immediately translates into the form of either the $T$ or
$\mathbf x_{\text{CM}}$ integrand, depending on the actual order of
integration. If the $T$ integral is done first, the resulting
loop-averaged $\mathbf x_{\text{CM}}$ integrand can be viewed as the
interaction energy density, the calculation of which is already an
instructive intermediate step.\footnote{The $\mathbf x_{\text{CM}}$
  integrand actually corresponds to a static effective-action density
  in the present case. The relation to the true interaction energy
  density which corresponds to the 00 component of the interaction
  energy-momentum tensor is give by a total derivative
  \cite{Schubert:2001he,Scardicchio:2005di}.} In principle, the $T$
and $\mathbf x_{\text{CM}}$ integration as well as the average over
all worldlines can be done in arbitrary order, depending on numerical
convenience.

There is, however, an important technical difference between taking
the worldline average before or after the integrations. The apparent
advantage of doing the worldline average first is that the resulting
$T$ and $\mathbf x_{{\text{CM}}}$ integrands are smooth, despite the
fact that the worldlines are fractal; this was exploited in many
worldline numerical applications so far. In the present work, we
nevertheless do the loop average at a later step. As a consequence,
the resulting integrands can become complicated in the sense that the
support of the integrand is a piecewise disconnected set. However,
once the support is determined by special algorithms, at least one
integral can be done analytically, since the integrand
$\sim\Theta_\Sigma=1$ and thus is extremely simple on the support.
This leads to significant numerical acceleration, constituting the
basic new ingredient of our improved algorithms.

\subsection{Sphere above plate}

\begin{figure}
\centering
\includegraphics[width=0.5\columnwidth]{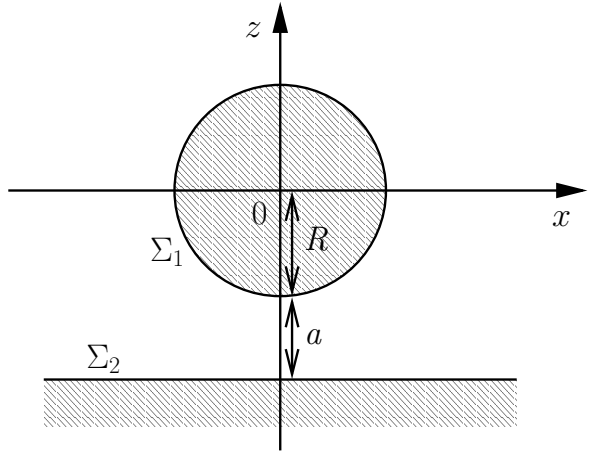}
\caption{Geometry of the sphere-plate configuration}
\label{fig:drawing1}
\end{figure}
The geometry of the sphere-plate configuration is illustrated in Fig.
\ref{fig:drawing1}. It is rotationally symmetric with respect to the
$z$ axis, therefore the three dimensional $x_\mathrm{CM}$ integration
in Eq. \eqref{eq:VEV2} trivially reduces to a two dimensional
integration.  We choose the following order of remaining
integrations/summations: first, we do the $T$ integral for each
worldline; then, we take the average over all loops, and finally
integrate over the resulting energy density. For the first step, the
numerically most challenging task is to determine the support
$\mathcal S_\ell$ of $\Theta_\Sigma[\mathbf x_\mathrm{CM}+\sqrt T\mathbf
y_\ell]$ on the $T$ axis to perform the $T$ integration in Eq.
(\ref{eq:ECasWd}). In the given geometry, $\Theta_\Sigma[\mathbf
x_\mathrm{CM}+\sqrt T\mathbf y_\ell]$
equals 1 if there exists a pair $k,l$, such that $\mathbf
x_\mathrm{CM}+\sqrt T\mathbf y_{\ell k}$ lies inside the sphere and
$\mathbf x_\mathrm{CM}+\sqrt T\mathbf y_{\ell l}$ lies below the plate;
otherwise it is zero.\footnote{Strictly speaking, this criterion
  misses the rare case that the link between two neighboring points
  which are both outside the sphere intersects the sphere. We neglect
  these contributions, since the verification of this pattern is much
  more time-consuming than simply increasing the amount of points per
  loop to reduce the corresponding systematic error.}  To investigate
the support $\mathcal S_\ell$, it turns out to be useful to distinguish
between $\mathbf x_\mathrm{CM}$ lying inside the sphere, $\mathbf
x_\mathrm{CM}^2\leq R^2$, and $\mathbf x_\mathrm{CM}$ lying outside,
$\mathbf x_\mathrm{CM}^2>R^2$, as the former case is much simpler than
the latter.

\subsubsection{Inside}

Inside the sphere, the support $\mathcal S_\ell$ is a single $T$
interval. The lower bound $T^{\mathrm{min}}_{\ell}$ is given by the $T$
value at which the loop $\mathbf x_\mathrm{CM}+\sqrt T\mathbf y_\ell$
touches the plate,
\begin{equation}
T^{\mathrm{min}}_{\ell}=\left(\frac{R+a+x_{\mathrm{CM}z}
}{\min_k y_{\ell kz}}\right)^2\label{eq:Tmin},
\end{equation}
where $\min_k y_{\ell kz}$ is the minimal $z$ coordinate of the unit
loop $\mathbf y_\ell$. The upper bound $T^{\mathrm{max}}_{\ell}$ is the
largest $T$ value for which the loop intersects the sphere,
\begin{align}
  T^{\mathrm{max}}_{\ell}&=\max\Bigl\{T:\exists_k
  \bigl(\mathbf x_\mathrm{CM}+\sqrt T\mathbf y_{\ell k}\bigr)^2=R^2\Bigr\}\\
  &=\max_k\left(-{\mathbf x_\mathrm{CM}\cdot\mathbf y_{\ell k} \over
      \mathbf y_{\ell k}^2}+ \sqrt{\left({\mathbf
          x_\mathrm{CM}\cdot\mathbf y_{\ell k} \over \mathbf y_{\ell
            k}^2}\right)^2 -{\mathbf x_\mathrm{CM}^2-R^2\over \mathbf
        y_{\ell k}^2}}\right)^2.
\end{align}
Performing the $T$ integration, we obtain the Casimir interaction
energy density $\varepsilon_\text{Casimir}$ inside the sphere,
$E_{\text{Casimir}}=\int
d^3x_{\text{CM}}\,\varepsilon_{\text{Casimir}}$,
\begin{equation}
\varepsilon_\text{Casimir}(\mathbf x_\mathrm{CM})={1\over64\pi^2}
{1\over n_\mathrm L}\sum_{\ell=1}^{n_\mathrm L}
\biggl(\frac{1}{(T^{\mathrm{max}}_{\ell})^2}
-\frac{1}{(T^{\mathrm{min}}_{\ell})^2}\biggr)
\cdot\theta(T^{\mathrm{max}}_{\ell}-T^{\mathrm{min}}_{\ell}),
\end{equation}
where the $\theta$ function takes care of the (non-generic) case that
the loop never intersects both surfaces.
\begin{figure}
\centering
\includegraphics[width=\columnwidth]{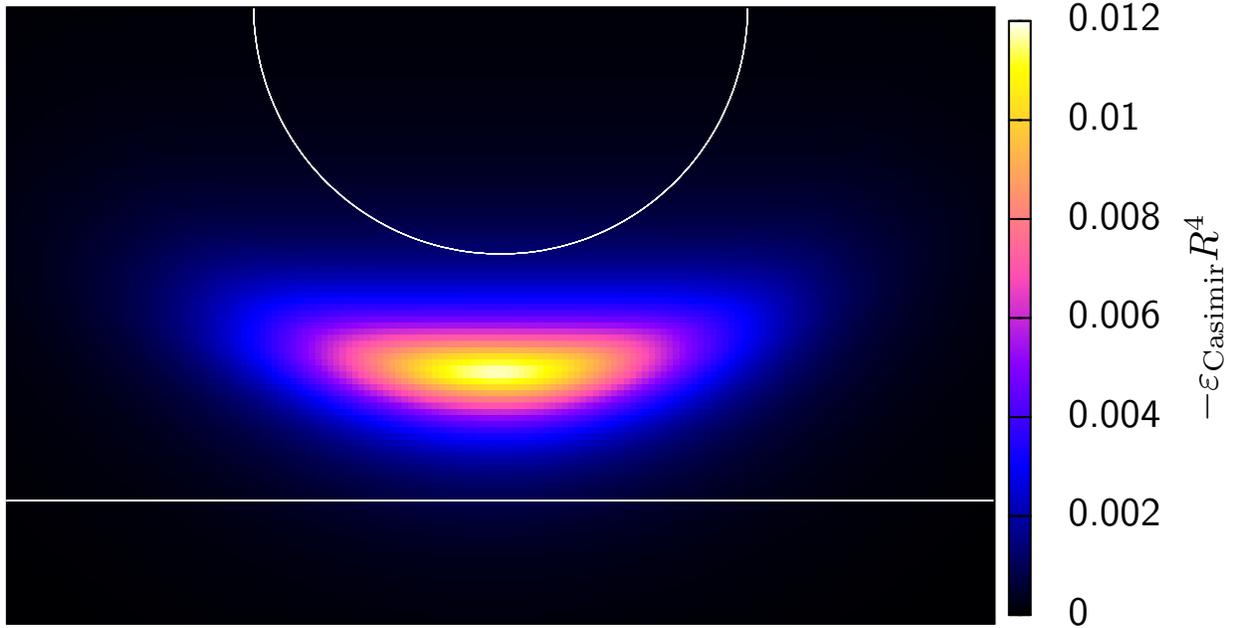}
\caption{Contour plot of the negative Casimir interaction energy
density $\varepsilon_\text{Casimir}$ for a sphere of radius $R$ above
an infinite plate; the sphere-plate separation $a$ has been chosen as
$a=R$ here.}
\label{fig:contour}
\end{figure}
This quantity is plotted in the contour plot Fig.~\ref{fig:contour}
in the region inside the white circle. The contribution to the total
Casimir interaction energy is small compared to the energy outside
the sphere. The density of the latter is shown in the same figure
outside the circle, obtained by the procedure described next.

\subsubsection{Outside}

\begin{figure}
\centering
\includegraphics[width=0.5\columnwidth]{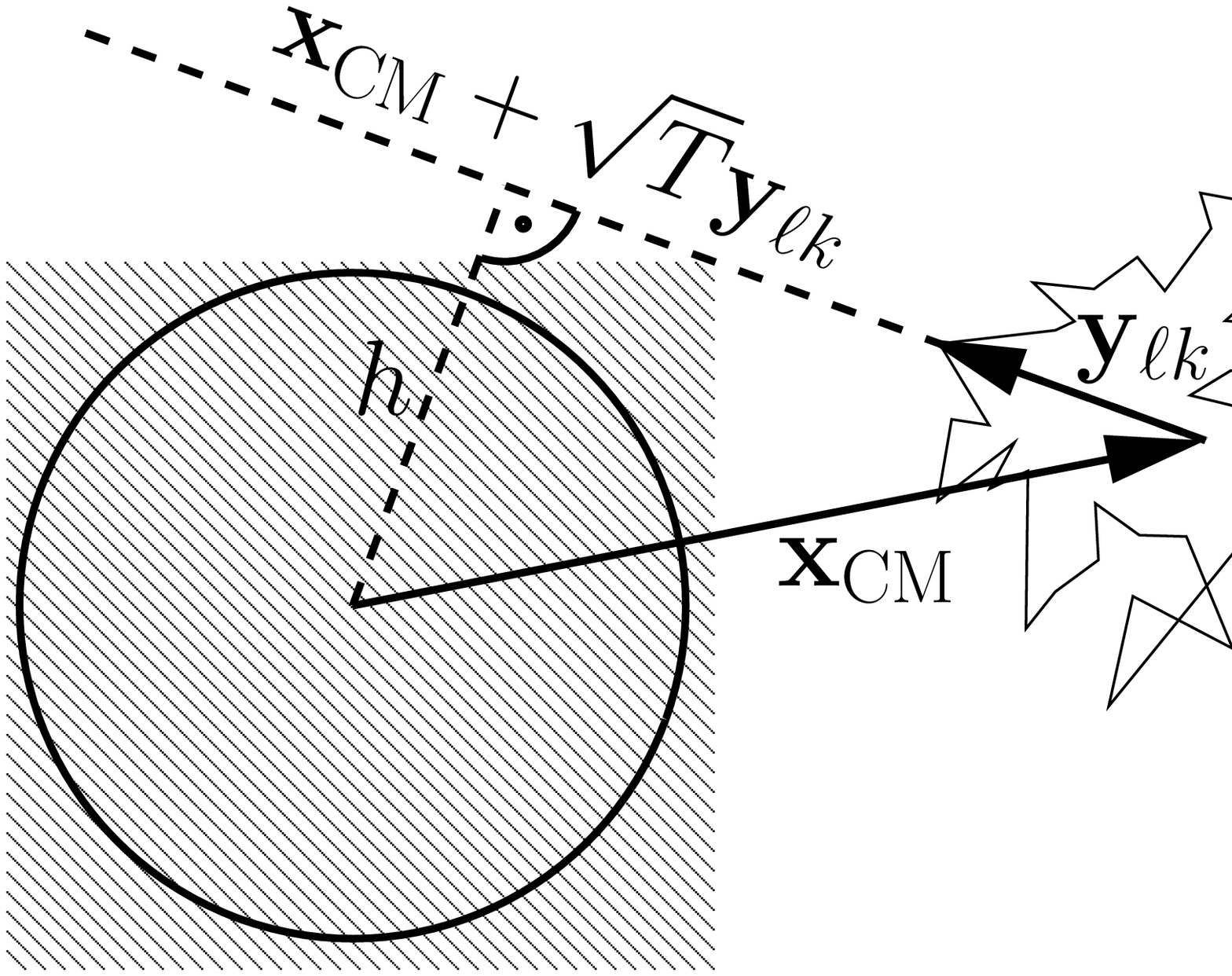}
\caption{For any propertime $T$, the ray
  $\mathbf x_\mathrm{CM}+\sqrt T\mathbf y_{\ell k}$ does not intersect
  the sphere. The corresponding point $\mathbf y_{\ell l}$ is thus
  not relevant for the interaction energy density at the given
  $\mathbf x_{\text{CM}}$ and consequently sorted out in a first step
  of the algorithm.}
\label{fig:drawing2}
\end{figure}

Outside the sphere the support $\mathcal S_\ell$ is not merely one
single interval as in the previous case, but a whole set of
successive intervals.
As illustrated in Fig.~\ref{fig:drawing2}, for a unit loop $\mathbf 
y_\ell$ at a center of mass $\mathbf x_\mathrm{CM}$ outside the sphere,
the ray $\mathbf x_\mathrm{CM}+\sqrt T\mathbf y_{\ell k}$ does not pierce
the sphere for most indices $k$. The corresponding points on the loop
are not relevant for the Casimir energy and the first step in our
algorithm is to sort them out. Two conditions are evaluated for this
purpose: a point $\mathbf y_{\ell k}$ is only relevant for further
computations if
\begin{enumerate}
\item the vector $\mathbf y_{\ell k}$ points towards the sphere, implying
\begin{equation}
\mathbf x_\mathrm{CM}\cdot\mathbf y_{\ell k}<0,
\end{equation}
\item the distance $h$ between the ray
$\mathbf x_\mathrm{CM}+\sqrt T\mathbf y_{\ell k}$
and the center of the sphere is smaller than the radius $R$,
\begin{equation}
h^2=\mathbf x_\mathrm{CM}^2-\left(\mathbf x_\mathrm{CM}\cdot
{\mathbf y_{\ell k}\over|\mathbf y_{\ell k}|}\right)^2
<R^2.
\end{equation}
\end{enumerate}
If these conditions are fulfilled, the $T$ values at which the ray
intersects the sphere are determined by
\begin{equation}
(\mathbf x_\mathrm{CM}+\sqrt T\mathbf y_{\ell k})^2=R^2,
\end{equation}
which has the solutions
\begin{equation}
T^\pm_{\ell k}=\left(-{\mathbf x_\mathrm{CM}\cdot\mathbf y_{\ell k}
\over \mathbf y_{\ell k}^2}
\pm\sqrt{\left({\mathbf x_\mathrm{CM}\cdot\mathbf y_{\ell k}
\over \mathbf y_{\ell k}^2}\right)^2
-{\mathbf x_\mathrm{CM}^2-R^2\over \mathbf y_{\ell k}^2}}\right)^2.
\end{equation}
For $T\in[T_{\ell k}^-,T_{\ell k}^+]$ the point $\mathbf x_\mathrm{CM}+\sqrt
T\mathbf y_{\ell k}$ lies inside the sphere and consequently we know that
the loop intersects the sphere. For a given loop, the total set of $T$
values for which this is the case is the union of the intervals
$[T_{\ell k}^-,T_{\ell k}^+]$ of all points $y_{\ell k}$ in the unit loop,
$\bigcup_k[T_{\ell k}^-,T_{\ell k}^+]$. Taking into account the minimal $T$
value for which the loop intersects the plate, $T^\mathrm{min}_\ell$, the
contribution of the unit loop to the propertime integrand has the
support
\begin{equation}
\mathcal S_\ell=[T^\mathrm{min}_\ell,\infty)
\cap\bigcup_k[T_{\ell k}^-,T_{\ell k}^+].\label{eq:supp}
\end{equation}
The set union can be determined efficiently by use of a sorting
algorithm like \emph{quicksort}, for example. Once $\mathcal S_\ell$ is
determined, the $T$ integration can be performed analytically. The
worldline estimate for the Casimir interaction energy density outside
the sphere therewith is
\begin{equation}
\varepsilon_\text{Casimir}(\mathbf x_\mathrm{CM})=-{1\over32\pi^2}
{1\over n_\mathrm L}\sum_{\ell=1}^{n_\mathrm L}
\int_{\mathcal S_\ell}{dT\over T^3},\label{eq:epsout}
\end{equation}
which is plotted in Fig. \ref{fig:contour} outside the white circle.

\subsubsection{Optimization}

The algorithm so far works well if the distance between sphere and
plate $a$ is of the same order of magnitude as the sphere's radius
$R$, $a\approx R$. To improve the accuracy by increasing the number
of loops $n_\mathrm L$ and the number of points per loop $N$, the
algorithm can be parallelized as \emph{embarrassingly parallel computation}
by dividing the loop ensemble into independently processed
sub-ensembles. However, if the two scales $a$ and $R$ differ
significantly in size, additional improvements of our
algorithm are advisable.\bigskip

\noindent
{\sf Large distances ($a\gg R$):} if the distance $a$ is large
compared to the radius $R$, the algorithm described so far becomes
inefficient due to the following reason: only loops with a minimal
extent of the same order of magnitude as the distance between
sphere and plate do contribute to the Casimir energy. For a loop
$\mathbf x_\mathrm{CM}+\sqrt T\mathbf y_\ell$ this means, that the unit
loop $\mathbf y_\ell$ has to be scaled by a large factor $\sqrt T$. This
implies that the distance between subsequent points on the loop
increases, too.  However, to ensure that the scaled loops still
resolve the sphere, this distance should be significantly smaller than
the sphere's radius.  Thus, the number of points per loop $N$ has to
be increased with increasing $a/R$.

A rough measure for the extent of a loop is the variance of the
coordinates of its points. The ensemble average of the variance for
large $N$ is $\langle (\sqrt T\mathbf y_{\ell k})^2\rangle=T/6$. As a consequence
we expect the $T$ integral to be dominated by
$T\approx6a^2$, also because the contribution
for large $T$ is damped by the $1/T^3$ factor. The root-mean-square of the
distance between two subsequent points on a loop for large
$N$ is $\sigma=\sqrt{2T/N}$. Using the dominating $T$ value we obtain
$\sigma\approx2\sqrt{3/N}a$. We demand this value to be much smaller than
the radius of the sphere, which implies $N\gg12a^2/R^2$. For a distance
$a=10R$, already much more than 1000 ppl have to be used, for
$a=100R$ much more than 100 000 ppl.

A slight modification enables our algorithm to cope with this high
resolution and the corresponding amount of data much more efficiently.
So far, for all center of masses $\mathbf x_\mathrm{CM}$ with a common
$z$ coordinate $\mathbf x_{\mathrm{CM}z}$, the first interval on the
right-hand side of Eq. \eqref{eq:supp}, $[T^\mathrm{min}_\ell,\infty)$,
is the same, which can be utilized to speed up the calculation. In
contrast, the union in the same equation is different for all centers of
masses.  However, it is this part of the equation which consumes most
of the CPU time. Modifying the transformation Eq. \eqref{eq:lt}
reverses the circumstances: let us define the rotation $R(\mathbf
x_\mathrm{CM},\mathbf e_z)$ by $\mathbf x_\mathrm{CM}/|\mathbf
x_\mathrm{CM}| =R(\mathbf x_\mathrm{CM},\mathbf e_z)\mathbf e_z$.  By
using
\begin{equation}
\mathbf x_{\ell k}=\mathbf x_\mathrm{CM}
+\sqrt TR(\mathbf x_\mathrm{CM},\mathbf e_z)\mathbf y_{\ell k}
\label{eq:rot}
\end{equation}
\begin{figure}
\centering
\includegraphics[width=0.5\columnwidth]{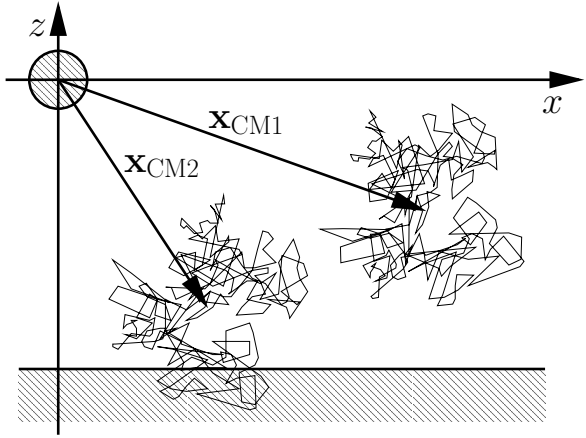}
\caption{Sketch of a unit loop at different centers of masses
$x_{\mathrm{CM}1}$ and $x_{\mathrm{CM}2}$, as used for large $a/R$.
The unit loop is rotated corresponding to the orientation of the
center of mass.}
\label{fig:rot}
\end{figure}
(see Fig. \ref{fig:rot}), the union in Eq. \eqref{eq:supp} is the same
for all center-of-mass values $x_\mathrm{CM}$ with a given absolute
value $|\mathbf x_\mathrm{CM}|$ and can be computed once and for all.
In turn, $T^\mathrm{min}_\ell$ is no longer degenerate with respect to
some $\mathbf x_{\text{CM}}$ coordinate.  The important advantage is
that this dependence can be computed much faster. For each loop, we
generate an array of its minimal $z$ coordinate as function of the
angle between $\mathbf x_\mathrm{CM}$ and $\mathbf e_z$.  The bound
$T^\mathrm{min}_\ell$ then results from Eq. \eqref{eq:Tmin}, where the
minimum is read from the array.
Note that the transformation \eqref{eq:rot} is a legitimate symmetry
operation for ensemble-averaged quantities, owing to the rotational
invariance of the exponential weight factor in Eq.  \eqref{eq:mean}.

There is a price to be paid for the desired feature of having a common
set union in Eq. \eqref{eq:supp} for different centers of masses
$x_\mathrm{CM}$: without the transformation, all points of a given
unit loop are equally involved in scanning the curvature of the
sphere. With the transformation, always the same points of a unit loop
are close to the sphere, independently of $\mathbf{x}_{\text{CM}}$.
This corresponds to a loss of statistics, implying a slight increase
of the statistical errors.  However, this is by far compensated for by
the gain in computation speed, which enables us to significantly
reduce again the statistical error by brute force.\bigskip

\noindent
{\sf Small distances ($a\ll R$):} the main contribution of the Casimir
interaction energy density is localized between sphere and plate. If
the distance $a$ is much smaller than the sphere's radius $R$, the
lower bound of the support $\mathcal S_\ell$ in that region is at very
small $T$ values compared to the upper bound of the support's first
interval. Since the $T$ integrand falls off rapidly with $1/T^3$, the
$T$ integral is dominated by this lower bound. Therefore, a very good
estimate is given by replacing $\mathcal S_\ell$ simply by the
interval $[T_\ell^{\text{min}},\infty)$, resulting in
\begin{equation}
\varepsilon_\text{Casimir}(x_\mathrm{CM})\simeq-{1\over64\pi^2}
{1\over n_\mathrm L}\sum_{\ell=1}^{n_\mathrm L}
{1\over (T_\ell^{\text{min}})^2}.\label{eq:sd}
\end{equation}
In particular outside the sphere, the numerical evaluation of this
expression is much faster than the evaluation of the full expression
Eq. \eqref{eq:epsout}.

For a rough estimate of the validity range of this approximation, we
use the ensemble's standard deviation of the point position, ${\langle
  (\sqrt T\mathbf y_{\ell k})^2\rangle}={T/6}$, to estimate the extent
of a loop. Right between sphere and plate, where the energy density is
largest, the lower bound of the propertime integral then is
approximately $T_\ell^{\text{min}}\approx 3a^2/2$. If the extent of the
loop increases beyond $a/2+2R$, we expect the loop to intersect the
sphere no longer for $T_1\gtrapprox6(a/2+2R)^2$. By setting the value
of $T_1$ to infinity instead, as done in Eq.~\eqref{eq:sd}, we
introduce a systematic error
$\Delta\varepsilon_\text{Casimir}/\varepsilon_\text{Casimir}
\lessapprox \bigl(1+4R/a\bigr)^{-4}$. For distances $a$ smaller than
$0.8R$ this error is smaller than one per mille. In this work, we have
used Eq.~\eqref{eq:sd} to compute high-precision values for $a<0.02R$.

\subsubsection{Results}

\begin{figure}[t]
\centering
\includegraphics[width=\columnwidth]{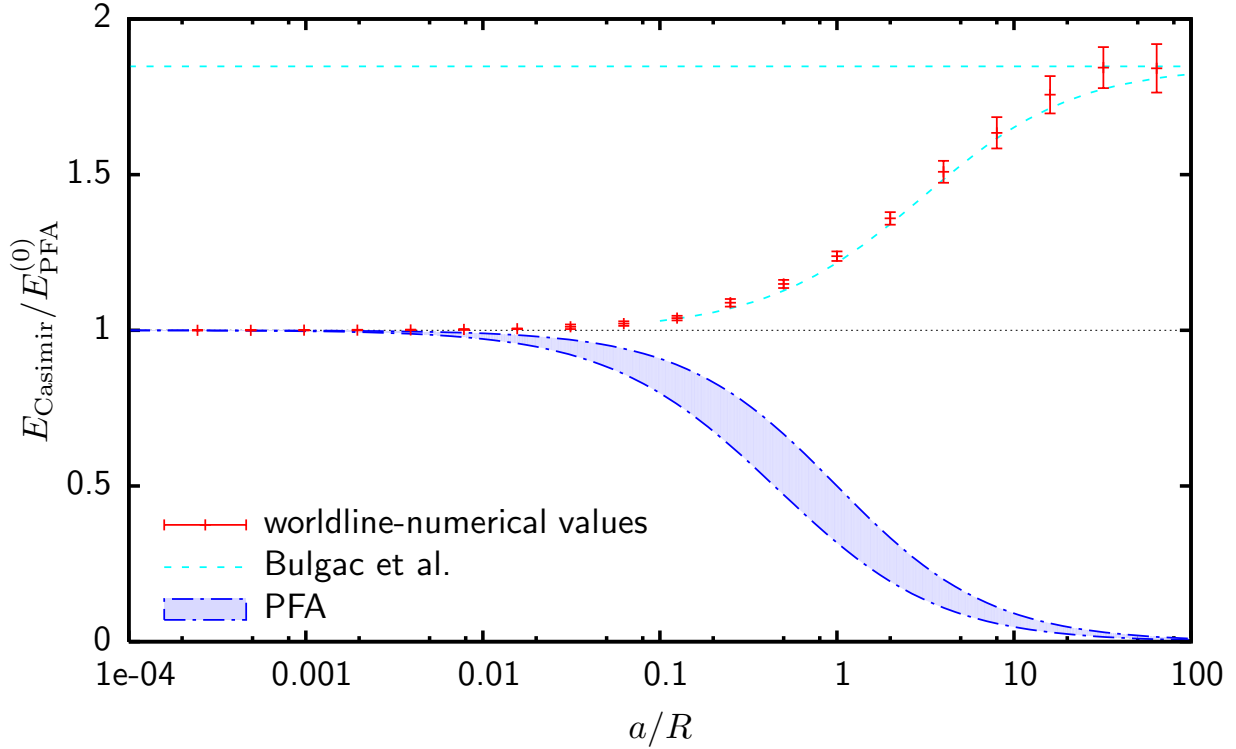}
\caption{Casimir interaction energy of a sphere with radius $R$ and an
infinite plate vs. the curvature parameter $a/R$. The energy is
normalized to the zeroth-order PFA formula (\ref{eq:Ezero}),
$E^{(0)}_{\text{PFA}}$. For larger curvature parameter, the PFA
estimate (dot-dashed line) differs qualitatively from the worldline
result (crosses with error bars). Here, we observe good agreement of
our result with the exact solution of \cite{Bulgac:2005ku} which is
available for $a/R\gtrsim 0.1$ (dashed line).}
\label{fig:sphere}
\end{figure}
Figure \ref{fig:sphere} presents a global view on the Casimir
interaction energy for a wide range of the curvature parameter $a/R$;
the energy is normalized to the zeroth order of the PFA formula
(\ref{eq:Ezero}), $E^{(0)}_{\text{PFA}}$.
For small $a/R$ (``large spheres''), our worldline result (crosses
with error bars) and the full sphere- and plate-based PFA estimates
(dashed-dotted lines) show reasonable agreement, settling at the
zeroth-order PFA $E^{(0)}_{\text{PFA}}$. The full PFA departs on
the percent level from $E^{(0)}_{\text{PFA}}$ for $a/R \gtrsim 0.01$,
exhibiting a relative energy decrease. By contrast, our worldline
result first stays close to $E^{(0)}_{\text{PFA}}$ and then increases
towards larger energy values relative to $E^{(0)}_{\text{PFA}}$. This
observation confirms earlier worldline studies \cite{Gies:2003cv} and
agrees with the optical approximation \cite{Scardicchio:2004fy} in
this curvature regime.

For larger curvature $a/R\gtrsim 0.1$ (``smaller spheres''),
we observe a strong increase relative to $E^{(0)}_{\text{PFA}}$
\cite{Gies:2005ym}. Here, our data satisfactorily agrees with the
exact solution found recently for this regime \cite{Bulgac:2005ku}
(dashed line). The latter work also provides for an exact asymptotic
limit for $a/R\to\infty$, resulting in $180/\pi^4$ for our
normalization.  Our worldline data confirms this limit in
Fig.  \ref{fig:sphere}.

Two important lessons can be learned from this plot: first, the PFA
already fails to predict the correct sign of the curvature effects
beyond zeroth order, see also \cite{Brevik:2004uw}. Second, the
relation between the Casimir effect for Dirichlet scalars and that for
the EM field is strongly geometry dependent. For the parallel-plate
case, Casimir forces only differ by the number of degrees of freedom,
cf. the coefficient $c_{\text{PP}}$ in Eq.~\eqref{eq:Enorm}. For large
curvature, the Casimir energy for the Dirichlet scalar scales with
$a^{-2}$, whereas that for the EM field obeys the Casimir-Polder law
$\sim a^{-4}$ \cite{Casimir:1947hx,DeKieviet}.  Already this
difference demonstrates that simple approximation methods such as the
PFA are highly problematic, since no reference to the nature of the
fluctuating field other than the coefficient $c_{\text{PP}}$ is made.

\begin{figure}
\centering
\includegraphics[width=\columnwidth]{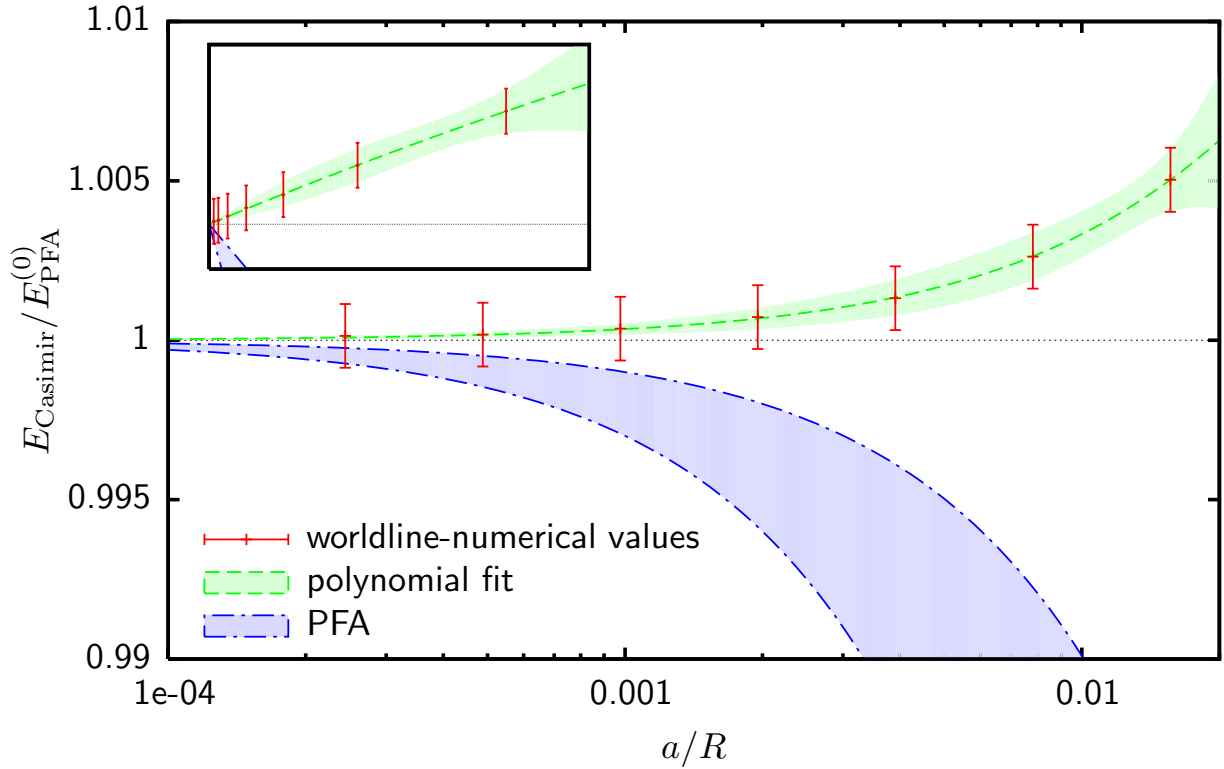}
\caption{Magnified view of Fig. \ref{fig:sphere} for small $a/R$. The
0.1\% validity range of the PFA is characterized by curvature
parameters, where the error band of our worldline results and the
PFA band (blue-shaded/in between the dot-dashed lines) overlap, see
Eq. \eqref{eq:acconepermille}. The dashed line depicts a constraint
polynomial fit of the worldline result,
$p(a/R)=1+0.35(a/R)-1.92(a/R)^2$, and its standard deviation, see
Eq.~\eqref{eq:spfit}. The inlay displays the same curves with a linear
$a/R$ axis. } 
\label{fig:sphere_small_a}
\end{figure}
%
%
For a quantitative determination of the PFA validity limits,
Fig. \ref{fig:sphere_small_a} displays the zeroth-order normalized
energy for small curvature parameter $a/R$. Here, our result has an
accuracy of 0.1\% (jack-knife analysis). The error is dominated by the
Monte Carlo sampling and the ordinary-integration
accuracy; the error from the worldline discretization is found
negligible in this regime, implying a sufficient proximity to the
continuum limit.

In addition to our numerical error band, we consider the region
between the sphere- and the plate-based PFA as the PFA error band.  We
identify the 0.1\% accuracy limit of the PFA with the curvature
parameter $a/R|_{0.1\%}$ where the two bands do no longer
overlap. We obtain
\begin{equation}
\frac{a}{R}\Big|^{\text{PFA}}_{0.1\%}\leq 0.00073
\label{eq:acconepermille}
\end{equation}
as the corresponding validity range for the curvature parameter.  For
instance, for a typical sphere with $R=200\mu$m and an experimental
accuracy goal of $0.1\%$, the PFA should not be used for $a\gtrsim
150$nm. We conclude that the PFA should be dropped from the analysis
of future experiments.

For the 1\% accuracy limit of the PFA, we increase the band of our
worldline estimate by this size and again determine the curvature
parameter for which there is no intersection with the PFA
band anymore. We obtain
\begin{equation}
\frac{a}{R}\Big|^{\text{PFA}}_{1\%}\leq 0.00755.
\label{eq:acconepercent} 
\end{equation}
For a sphere with $R=200\mu$m and an experimental accuracy goal of
$1\%$, the PFA holds for $a<1.5\mu$m. This result confirms the use of
the PFA for the data analysis of the corresponding experiments
performed so far.  

In order to study the asymptotic expansion of the normalized energy,
we fit our worldline numerical data to a second-order polynomial for
$a/R<0.1$ and include the exactly known result for $a/R\to 0$. We
obtain
\begin{equation}
E^{\text{sphere-plate}}_{\text{WN data fit}}
=-c_{\text{PP}}\frac{\pi^3}{1440}
\frac{R}{a^2}\left(1+ 0.35
\frac{a}{R} -1.92 \frac{a^2}{R^2} \pm 0.19 \frac{a}{R}
 \sqrt{ 1 - 137.2 \frac{a}{R} + 5125 \frac{a^2}{R^2}}\right),
 \label{eq:spfit} 
\end{equation}
valid for $a/R<0.1$; here, $c_{\text{PP}}=1$ for the real and
$c_{\text{PP}}=2$ for a complex Dirichlet scalar. The fit result is
plotted in Fig.~\ref{fig:sphere_small_a} (dashed lines), which
illustrates that $E\simeq E_{\text{WN data fit}}$ is a satisfactory
approximation to the Casimir energy for $a/R<0.1$, replacing the PFA
\eqref{eq:Enorm}.  The inlay of Figure \ref{fig:sphere_small_a}
displays the same curves with a linear $a/R$ axis, illustrating that
the lowest-order curvature effect is linear in $a/R$.
A more direct result for the linear curvature coefficient can be
obtained by a constraint linear fit; in this simpler case, the fit
polynomial yields $p_{\text{fit}}(a/R)=1+(0.33\pm0.06) \frac{a}{R}$
instead of the expression in parentheses in Eq.~\eqref{eq:spfit}.
Given the results of the PFA \eqref{eq:Enorm}, the semiclassical
approximation \cite{semicl}, $p_{sc}(a/R)\simeq1-0.17\frac{a}{R}$, cf.
\cite{Bulgac:2005ku}, and the optical approximation
\cite{Scardicchio:2004fy}, $p_{opt}(a/R)\simeq1+0.05\frac{a}{R}$, the
latter appears to estimate curvature effects more appropriately; but
all these  approximations are not quantitatively reliable for
beyond-zeroth-order curvature effects.

\subsection{Cylinder above plate}

\begin{figure}[t]
\centering
\includegraphics[width=\columnwidth]{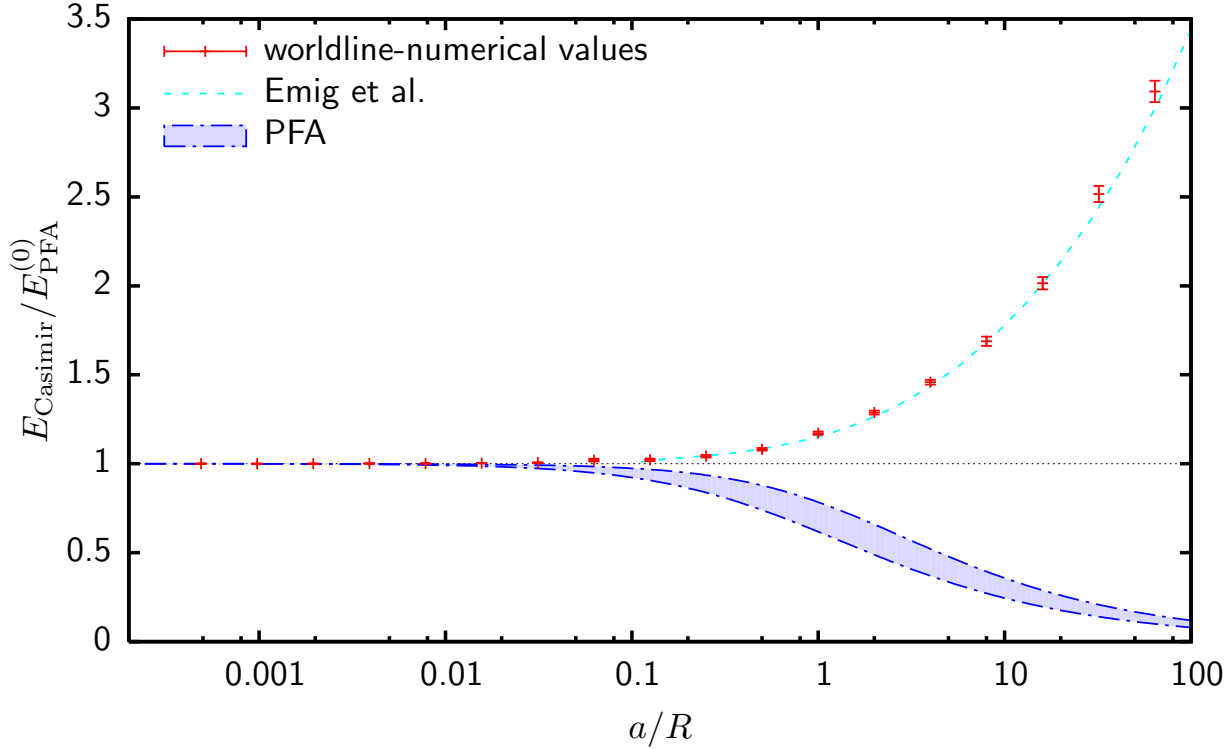}
\caption{Casimir interaction energy (normalized to
  $E_{\text{PFA}}^{(0)}$) of an infinitely long cylinder with radius
  $R$ at a distance $a$ above an infinite plate vs. the curvature
  parameter $a/R$.  We observe good agreement of our result with
  the exact solution of \cite{Emig:2006uh} which is available for
  $a/R\gtrsim 0.1$ (dashed line).}
\label{fig:cylinder}
\end{figure}

\begin{figure}
\centering 
\includegraphics[width=\columnwidth]{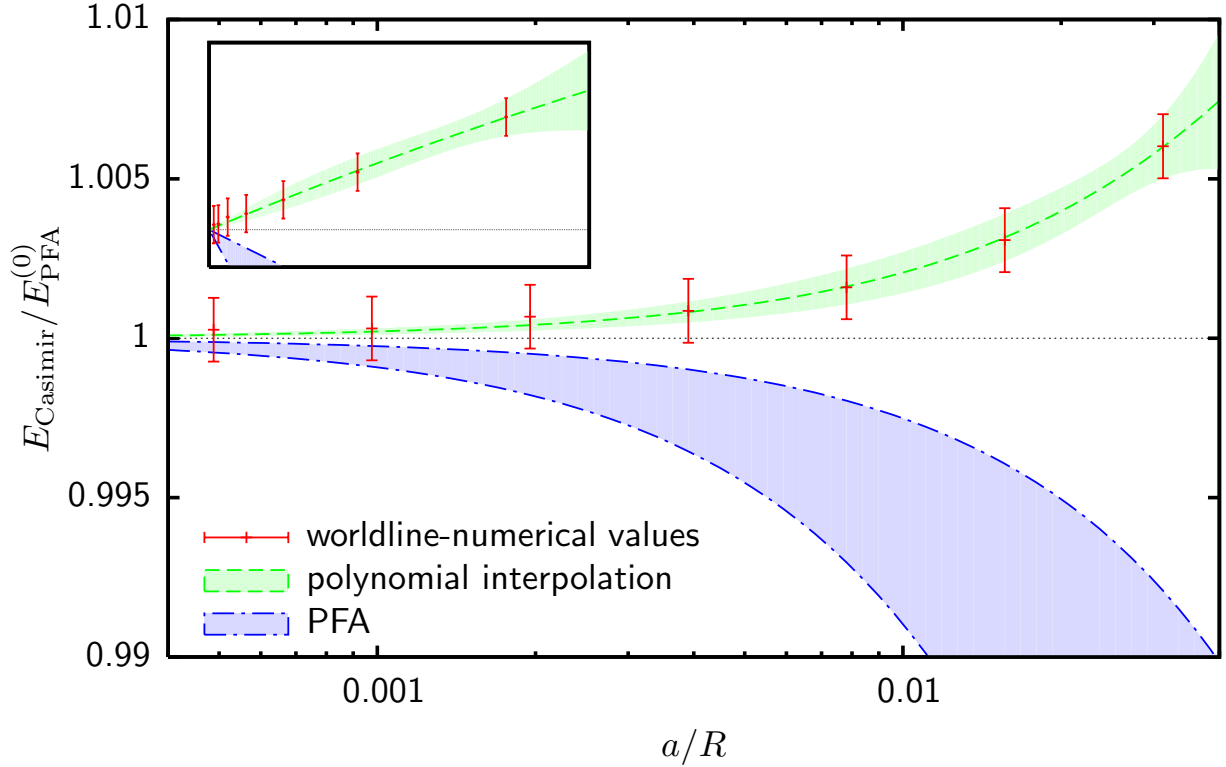}
\caption{Magnified view of Fig.~\ref{fig:cylinder}
for small values of $a/R$. The dashed line with error band depicts a
constraint polynomial fit to the numerical data,
$p(a/R)=1+0.21(a/R)-0.66(a/R)^2$, and its standard deviation.
 The inlay displays the same curves with a linear
$a/R$ axis.}
\label{fig:cylinder_small_a}
\end{figure}

The cylinder-plate configuration is a promising tool for
high-precision experiments \cite{Brown-Hayes:2005uf}, since the force
signal increases linearly with the cylinder length. The numerics is
very similar to the sphere-plate configuration, even less
computing power is required, because only two dimensional loops have
to be processed due to the translational symmetry.  Figure
\ref{fig:cylinder} shows the corresponding Casimir interaction energy
versus the curvature parameter. The energy axis is again normalized to
the zeroth-order PFA result, 
\begin{equation}
E_{\text{PFA}}^{(0)}(a,R)=
-c_{\text{PP}} \frac{\pi^3}{1920\sqrt{2}}\,\,
\frac{R^{1/2}}{a^{5/2}}.
\end{equation}
A magnified view of the small curvature region in
Fig.~\ref{fig:cylinder} is shown in Fig.~\ref{fig:cylinder_small_a}.
As for the sphere-plate configuration, we fit our data to a
second-order polynomial in this range, including the exactly known
result for $a/R\to 0$, yielding
\begin{equation}
E^{\text{cylinder-plate}}_{\text{WN data fit}}\!
=-\frac{c_{\text{PP}}\pi^3}{1920\sqrt{2}}\, 
\frac{R^{1/2}}{a^{5/2}}
\left(\!1+ 0.21
\frac{a}{R} -0.66 \frac{a^2}{R^2} \pm 0.097 \frac{a}{R}
 \sqrt{ 1 - 68.60 \frac{a}{R} + 1282 \frac{a^2}{R^2}}\right),
\label{eq:cpfit} 
\end{equation}
for ${a}/{R}<0.1$.  The inlay of Fig.~\ref{fig:cylinder_small_a} shows
the same data with a linear $a/R$ axis. As for the sphere-plate
geometry, this plot demonstrates that the lowest-order curvature
effect is linear in $a/R$. A simpler linear fit to our data results in
$p_{\text{fit}}(x)\simeq1+(0.195\pm0.028)\frac{a}{R}$. This is in
remarkable agreement with the recently found analytical result
$p(a/R)=1+0.19\bar4 \frac{a}{R}+\mathcal O(a^2/R^2)$
\cite{Bordag:2006vc}, which represents a strong confirmation for both
methods.

The qualitative conclusions for the validity of the PFA are similar to
that for the sphere above a plate: beyond leading order, the PFA even
predicts the wrong sign of the curvature effects. Quantitatively, the
PFA validity limits are a factor $\sim3$ larger than Eqs.
\eqref{eq:acconepermille},\eqref{eq:acconepercent}, owing to the
absence of curvature along the cylinder axis.

The most important difference to the sphere-plate case arises for
large $a/R$. Here, the data is compatible with a log-like increase
relative to $E_{\text{PFA}}^{(0)}$, implying a surprisingly weak
decrease of the Casimir force for large curvature $a/R\to \infty$.
Our result agrees nicely with the recent exact result \cite{Emig:2006uh}
which is available for $a/R\gtrsim0.1$. The data thus confirms the
observation of \cite{Emig:2006uh} that the resulting Casimir force has
the weakest possible decay, $F\sim 1/[a^3\ln(a/R)]$, for
asymptotically large curvature parameter $a/R\to \infty$.

%

\subsection{Parallel plates revisited}

\begin{figure}
\centering
\includegraphics[width=0.5\columnwidth]{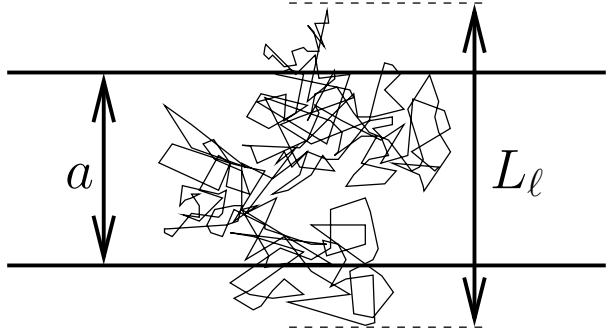}
\caption{Geometry of Casimir's parallel-plate configuration. A loop
  contributes to the Casimir interaction energy if its extent $L_\ell$
  along the $z$ direction is larger than the plate distance $a$.}
\label{fig:drawing4}
\end{figure}
As discussed at the beginning of this section, the order of the $T$
and $\mathbf x_{\text{CM}}$ integration and the ensemble averaging can
be chosen arbitrarily. As an example for an ``unusual'' order, let us
reconsider Casimir's classic parallel-plate configuration in $D=d+1$
dimensional spacetime, doing the $\mathbf x_{\text{CM}}$ integral
first and keeping the ensemble average till the very end. This will
reveal an unexpected mapping between the $D$-dimensional Casimir
effect and standard polymer physics.

In $d$ space dimensions, the surface or area volume $A$ of the Casimir
plates is taken as $d-1$ dimensional. The two (hyper-)plates are
separated by a distance $a$ along the $z$ direction which is normal to
the plates, see Fig.~\ref{fig:drawing4}. For this configuration, the
Casimir interaction energy for the massless Dirichlet scalar boils
down to
\begin{equation}
E_{\text{Casimir}}=-A\, \frac{1}{2(4\pi)^{D/2}}
\frac{1}{n_{\text{L}}} \sum_{\ell=1}^{n_{\text{L}}}\int_0^\infty
\frac{dT}{T^{1+D/2}}\, \int^\infty_{-\infty}\, dz_{\text{CM}}\,
  \Theta[z_{\text{CM}}+\sqrt{T} y_{z\ell}],
\label{eq:DCas1}
\end{equation}
where $y_{z\ell}$ denotes the $z$ coordinate of the $\ell$th unit
loop. Let us denote the extent of the $\ell$th unit loop in the $z$
direction by $L_\ell$,
\begin{equation}
L_\ell:=\max_{i,\,k} \left( |y_{z\ell i} - y_{z\ell k}| \right), \label{eq:DCas2}
\end{equation}
see Fig.~\ref{fig:drawing4}.  A scaled unit loop intersects both
plates if $\sqrt{T} L_\ell\geq a$. For a given unit loop with extent
$L_\ell$ and for a given propertime value $T$, the support of the
$z_{\text{CM}}$ integral corresponds to an interval
$I_z(T,L_\ell)=\sqrt{T}L_\ell - a$. Independently of the precise location of
this interval on the $z_{\text{CM}}$ axis, the $z_{\text{CM}}$
integral yields,
\begin{equation}
 \int_{-\infty}^\infty\, dz_{\text{CM}}\,
  \Theta[z_{\text{CM}}+\sqrt{T} y_{z\ell}]
=(\sqrt{T} L_\ell - a)\, \theta (\sqrt{T} L_\ell -a). \label{eq:DCas3}
\end{equation}
Now, also the $T$ integral can be done analytically, resulting in
\begin{eqnarray}
E_{\text{Casimir}}&=&-\frac{A}{a^{D-1}}\, \frac{1}{D(D-1)(4\pi)^{D/2}} 
\frac{1}{n_{\text{L}}} \sum_{\ell=1}^{n_{\text{L}}} L_\ell^{D} \nonumber\\
&=&-\frac{A}{a^{D-1}}\, \frac{1}{D(D-1)(4\pi)^{D/2}} 
\langle L_\ell^{D}\, \rangle. \label{eq:DCas4}
\end{eqnarray}
We observe that the Casimir interaction energy of the parallel-plate
configuration in $D=d+1$ spacetime dimensions is proportional to the
$D$th moment of the ensemble-averaged extent of a unit loop. This
ensemble average could easily be performed, which would lead us back to
the results of \cite{Gies:2003cv}.

Here, we will be satisfied by hilighting Eq.~\eqref{eq:DCas4} from a
different perspective, namely, in the language of
polymer physics. The Gau\ss ian velocity distribution of our
worldlines is identical to the Hamiltonian of a polymer, i.e., the
continuum limit of a random chain, without self-avoidance or
excluded-volume effect \cite{Kleinert:2004ev} in the limit of zero
end-to-end distance. In this language, $L_\ell$ corresponds to the
maximum spatial extent of the closed polymer measured in units of
$\sqrt{l_{\text{p}} c/(2D_{\text{p}})}$. Here, $l_{\text{p}}=N c$
denotes the total length of the polymer, $c$ is the chain length, and
$D_{\text{p}}$ is the number of dimensions in which the polymer can
move; the latter is completely arbitrary and independent of the
dimensionality of the Casimir system. Since $L_\ell$ is a highly
nonlocal object, its ensemble average is actually not so easily
computable by standard methods. Our result Eq.~\eqref{eq:DCas4} now
maps the problem of computing any $D$th moment of $L_\ell$ on the
$D$-dimensional Casimir problem for parallel plates. Using the
standard result for the latter \cite{Verschelde:1985jr},
\begin{equation}
E_{\text{Casimir}}=-\frac{A}{a^{D-1}}\, \frac{1}{(4\pi)^{D/2}}\,
  \Gamma(D/2)\, \zeta(D), \label{eq:DCas5}
\end{equation}
we obtain by comparison with Eq.~\eqref{eq:DCas4},
\begin{equation}
\langle L_\ell^D[y]\rangle =D(D-1)\, \Gamma(D/2)\, \zeta(D),
\end{equation}
a result that we have so far not been able to find in the literature
of polymer physics. Even the limit $D\to 1$ can be taken,
corresponding to the Casimir effect in zero space dimensions: this
results in $\langle L_\ell[y]\rangle =\sqrt{\pi}$ for the
average extent of a closed polymer.

\section{Conclusions}
\label{sec:C}

We have presented improved worldline numerical algorithms that can
efficiently deal with Casimir configurations involving curved
surfaces. We have used these algorithms to compute Casimir interaction
energies for the sphere-plate and cylinder-plate configuration induced
by a scalar field with Dirichlet boundary conditions. These
computations are done from first principles for a wide range of
curvature parameters $a/R$.  In general, we observe that curvature
effects and geometry dependencies are intriguingly rich, implying that
naive estimates can easily be misguiding. In particular, predictions
based on the PFA are only reliable in the asymptotic no-curvature
limit with quantitative validity bounds given above. We have
constructed polynomial fits of our results which can be used in the
small-curvature regime, $a/R\lesssim 0.1$, as a well-founded
substitute for the PFA formulas. Given the size of the true curvature
corrections for the Dirichlet scalar, we expect that genuine Casimir
curvature effects are in reach of currently planned experiments. In
this spirit, the so-called lateral Casimir force for corrugated
surfaces has recently been proposed as a suitable candidate for
identifying non-trivial geometry dependences beyond the PFA
\cite{Rodrigues:2006ku}.

Beyond the Dirichlet scalar investigated here, it is well possible,
e.g., for the EM field, that some cancellation of curvature effects
occurs between modes obeying different boundary conditions. In fact,
such a partial cancellation between TE and TM modes of the separable
cylinder-plate geometry can be observed in the recent exact result for
the EM field for medium curvature \cite{Emig:2006uh}; for small
curvature, curvature effects can even reverse sign
\cite{Bordag:2006vc}. More quantitatively, the TM mode in the
cylinder-plate case obeys Dirichlet boundary condition and thus
contributes, e.g., to the linear curvature correction with a
coefficient $\simeq 0.194$, as discussed below Eq.~\eqref{eq:cpfit};
the TE mode obeys Neumann boundary conditions, giving a negative
contribution which in total turns this linear coefficient for the EM
field into $\simeq -0.48$ \cite{Bordag:2006vc}. The latter result, in
fact, lies in the broad range of $[-0.92,-0.25]$ spanned by the PFA;
since the PFA does not make any reference to the nature of the
fluctuating field, this rough coincidence is, of course, purely
accidental. This strong dependence of Casimir curvature effects on the
nature of the fluctuating fields alone demonstrates already that
approximations ignoring this difference such as the PFA cannot be
trusted.  We emphasize again that Casimir calculations for the EM
field in non-separable geometries, such as the important sphere-plate
case, remain a prominent open problem. 

From a technical point of view, we would like to stress that our
results demonstrate the capability of worldline numerics for
performing high-precision computations with comparatively little
computing power. The simple scalability of the algorithms and the
flexibility for adapting them to arbitrary geometries makes worldline
numerics a unique tool for computing quantum energies.

Our algorithmic strategies also revealed an unexpected mapping between
the $D$-dimen\-sional parallel-plate Casimir effect and aspects of a
random-chain polymer ensemble. The origin of this mapping, of course,
lies in the fact that both quantum fluctuations in Casimir systems as
well as a polymer ensemble can be described by Feynman path integrals.
In the present case, the mapping can be utilized to transform a
comparatively difficult polymer problem into a field-theoretic Casimir
problem which can be solved by a variety of techniques. We believe
that this mapping is just a special case of a more general class of
mappings with potentially fruitful applications in both directions.

\section*{Acknowledgment}

The authors are grateful to T.~Emig, R.L.~Jaffe, A.~Scardicchio, and
A.~Wirzba for useful discussions and W.~Wetzel for providing the
expertise for parallelizing our algorithms.  The authors acknowledge
support by the DFG under contract Gi 328/1-3 (Emmy-Noether program)
and Gi 328/3-2.


\begin{thebibliography}{99}
\setlength{\itemsep}{-0.1mm} 
{\small
\parskip=0pt


\bibitem{Casimir:dh}
H.B.G.~Casimir,
Kon.\ Ned.\ Akad.\ Wetensch.\ Proc.\  {\bf 51}, 793 (1948).

\bibitem{Lamoreaux:1996wh}
S.~K.~Lamoreaux,
Phys.\ Rev.\ Lett.\  {\bf 78}, 5 (1997).

\bibitem{Mohideen:1998iz}
U.~Mohideen and A.~Roy,
Phys.\ Rev.\ Lett.\  {\bf 81}, 4549 (1998);

\bibitem{Roy:1999dx}
A.~Roy, C.~Y.~Lin and U.~Mohideen,
Phys.\ Rev.\ D {\bf 60}, 111101 (1999).

\bibitem{Ederth:2000}
T.~Ederth, Phys.\ Rev.\  A {\bf 62}, 062104 (2000)

\bibitem{Chan:2001}
H.B.~Chan, V.A.~Aksyuk, R.N.~Kleiman, D.J.~Bishop and F.~Capasso,
Science 291, 1941 (2001).

\bibitem{Chen:2002}
F.~Chen, U.~Mohideen, G.L.~Klimchitskaya and V.M.~Mostepanenko,
Phys.\ Rev.\ Lett.\ {\bf 88}, 101801 (2002).

\bibitem{Bordag:1998nv}
M.~Bordag, B.~Geyer, G.~L.~Klimchitskaya and V.~M.~Mostepanenko,
Phys.\ Rev.\ D {\bf 58}, 075003 (1998), 
{\em ibid.} {\bf 60}, 055004 (1999), 
{\em ibid.} {\bf 62}, 011701 (2000).

\bibitem{Long:1998dk}
J.~C.~Long, H.~W.~Chan and J.~C.~Price,
Nucl.\ Phys.\ B {\bf 539}, 23 (1999).


\bibitem{Mostepanenko:2001fx}
V.~M.~Mostepanenko and M.~Novello,
Phys.\ Rev.\ D {\bf 63}, 115003 (2001).


\bibitem{Milton:2001np}
K.~A.~Milton, R.~Kantowski, C.~Kao and Y.~Wang,
Mod.\ Phys.\ Lett.\ A {\bf 16}, 2281 (2001).


\bibitem{Decca:2003td}
R.S.~Decca, E.~Fischbach, G.L.~Klimchitskaya, D.E.~Krause, D.L.~Lopez and V.M.~Mostepanenko,
Phys.\ Rev.\ D {\bf 68}, 116003 (2003), 
R.S.~Decca, D.~Lopez, H.B.~Chan, E.~Fischbach, D.E. Krause and C.R.~Jamell,
Phys.\ Rev.\ Lett.\  {\bf 94}, 240401 (2005).

\bibitem{Klimtchiskaya:1999}
G.L.~Klimchitskaya, A.~Roy, U.~Mohideen, and V.M.~Mostepanenko, Phys.\
Rev.\ A {\bf 60}, 3487 (1999).

\bibitem{Lambrecht:1999vd}
A.~Lambrecht and S.~Reynaud,
Eur.\ Phys.\ J.\ D {\bf 8}, 309 (2000).

\bibitem{Bezerra:2000} 
V.B.~Bezerra, G.L.~Klimchitskaya, and V.M.~Mostepanenko,
Phys. Rev. A 62, 014102 (2000).

\bibitem{Bordag:2001qi}
M.~Bordag, U.~Mohideen and V.~M.~Mostepanenko,
Phys.\ Rept.\  {\bf 353}, 1 (2001).

\bibitem{Lambrecht:2005} 
P.A.~Maia Neto, A.~Lambrecht, and S.~Reynaud, Europhys.~Lett.~{\bf
 69}, 924 (2005); Phys.~Rev. A {\bf 72}, 012115 (2005).

\bibitem{Mostepanenko:2005qh}
V.~M.~Mostepanenko {\it et al.},
arXiv:quant-ph/0512134.

\bibitem{Brevik:2006jw}
I.~Brevik, S.~A.~Ellingsen and K.~A.~Milton,
arXiv:quant-ph/0605005.

\bibitem{Sernelius}
M.~Bostr\"om and Bo E.~Sernelius, Phys.~Rev.~Lett.~{\bf 84}, 4757 (2000).

\bibitem{Graham:2002xq}
N.~Graham, R.~L.~Jaffe, V.~Khemani, M.~Quandt, M. Scandurra and H.~Weigel,
Nucl.\ Phys.\ B {\bf 645}, 49 (2002).

\bibitem{Bressi:2002fr}
G.~Bressi, G.~Carugno, R.~Onofrio and G.~Ruoso,
Phys.\ Rev.\ Lett.\  {\bf 88}, 041804 (2002).
 
\bibitem{pft1}
B.V.~Derjaguin, I.I.~Abrikosova, E.M.~Lifshitz, Q.Rev. {\bf 10}, 295
(1956); 
J.~Blocki, J.~Randrup, W.J.~Swiatecki, C.F.~Tsang, Ann.~Phys.~(N.Y.)
{\bf 105}, 427 (1977).


\bibitem{semicl}
M. Schaden and L. Spruch, Phys.~Rev.~A {\bf 58}, 935 (1998);
Phys. Rev. Lett. {\bf 84} 459 (2000) 

\bibitem{Golestanian:1998bx}
R.~Golestanian and M.~Kardar, Phys.~Rev.~A {\bf 58}, 1713 (1998);
T.~Emig, A.~Hanke and M.~Kardar,
Phys.\ Rev.\ Lett.\  {\bf 87} (2001) 260402;
T.~Emig and R.~Buscher,
Nucl.\ Phys.\ B {\bf 696}, 468 (2004).

\bibitem{Gies:2003cv}
H.~Gies, K.~Langfeld and L.~Moyaerts,
JHEP {\bf 0306}, 018 (2003); 
arXiv:hep-th/0311168.

\bibitem{Scardicchio:2004fy}
A.~Scardicchio and R.~L.~Jaffe,
Nucl.\ Phys.\ B {\bf 704}, 552 (2005);
Phys.\ Rev.\ Lett.\  {\bf 92}, 070402 (2004).

\bibitem{Bulgac:2005ku}
A.~Bulgac, P.~Magierski and A.~Wirzba,
Phys.\ Rev.\ D {\bf 73}, 025007 (2006)
[arXiv:hep-th/0511056];
A.~Wirzba, A.~Bulgac and P.~Magierski,
J.\ Phys.\ A {\bf 39} (2006) 6815
[arXiv:quant-ph/0511057].

\bibitem{Emig:2006uh}
T.~Emig, R.~L.~Jaffe, M.~Kardar and A.~Scardicchio,
Phys.\ Rev.\ Lett.\  {\bf 96} (2006) 080403
[arXiv:cond-mat/0601055].

\bibitem{Bordag:2006vc}
M.~Bordag,
arXiv:hep-th/0602295.


\bibitem{Gies:2001zp}
H.~Gies and K.~Langfeld,
Nucl.\ Phys.\ B {\bf 613}, 353 (2001); 
Int.\ J.\ Mod.\ Phys.\ A {\bf 17}, 966 (2002).

\bibitem{Schubert:2001he}
see, e.g., C.~Schubert,
Phys.\ Rept.\  {\bf 355}, 73 (2001).

\bibitem{Gies:2005ym}
H.~Gies and K.~Klingmuller,
 J.~Phys.~A {\bf 39} 6415 (2006) [arXiv:hep-th/0511092].

\bibitem{Gies:2006bt}
H.~Gies and K.~Klingmuller,
arXiv:quant-ph/0601094, to appear in Phys.~Rev.~Lett. (2006).

\bibitem{Deutsch:1978sc}
D.~Deutsch and P.~Candelas,
Phys.\ Rev.\ D {\bf 20}, 3063 (1979); 
%
P.~Candelas,
Annals Phys.\  {\bf 143}, 241 (1982).

\bibitem{Barton:2001wd}
G.~Barton,
J.\ Phys.\ A {\bf 34}, 4083 (2001).

\bibitem{Graham:2003ib}
N.~Graham, R.~L.~Jaffe, V.~Khemani, M.~Quandt, O.~Schroeder and H.~Weigel,
Nucl.\ Phys.\ B {\bf 677}, 379 (2004)
[arXiv:hep-th/0309130]; 
%
H.~Weigel,
arXiv:hep-th/0310301.

\bibitem{Kenneth:2006vr}
O.~Kenneth and I.~Klich,
arXiv:quant-ph/0601011.

\bibitem{Scardicchio:2005di}
A.~Scardicchio and R.~L.~Jaffe,
Nucl.\ Phys.\ B {\bf 743} (2006) 249
[arXiv:quant-ph/0507042].

\bibitem{Brevik:2004uw}
I.~Brevik, E.K.~Dahl and G.O.~Myhr,
J.\ Phys.\ A {\bf 38}, L49 (2005).

\bibitem{Casimir:1947hx}
H.B.G.~Casimir and D.~Polder,
Phys.\ Rev.\  {\bf 73}, 360 (1948).

\bibitem{DeKieviet}
V.~Druzhinina and M.~DeKieviet, Phys.~Rev.~Lett.~ {\bf  91}, 193202
(2003). 

\bibitem{Brown-Hayes:2005uf}
M.~Brown-Hayes, D.A.R.~Dalvit, F.D.~Mazzitelli, W.J. Kim and R.~Onofrio,
Phys.\ Rev.\ A {\bf 72}, 052102 (2005).

\bibitem{Kleinert:2004ev}
H.~Kleinert,
``{PATH INTEGRALS} in Quantum Mechanics,  Statistics, Polymer Physics, and
Financial Markets ,''
 World Scientific, Singapore (2004).

\bibitem{Verschelde:1985jr}
H.~Verschelde, L.~Wille and P.~Phariseau,
Phys.\ Lett.\ B {\bf 149}, 396 (1984); 
N.~F.~Svaiter and B.~F.~Svaiter,
J.\ Math.\ Phys.\  {\bf 32}, 175 (1991).

\bibitem{Rodrigues:2006ku}
R.~B.~Rodrigues, P.~A.~Maia Neto, A.~Lambrecht and S.~Reynaud,
Phys.\ Rev.\ Lett.\  {\bf 96}, 100402 (2006)
[arXiv:quant-ph/0603120].
}
\end{thebibliography}
\end{document}